\documentclass[aps,prl,twocolumn,floats,showpacs,superscriptaddress]{revtex4-1}

\usepackage{bm}
\usepackage{graphicx}

\def \Grenoble{Laboratoire National des Champs Magn\'etiques Intenses, CNRS-UJF-UPS-INSA, 38042 Grenoble, France}
\def \WHOZA{Institute of Experimental Physics, University of Warsaw, ul. Ho\.za 69, 00-681 Warszawa, Poland}
\def \WPAC{Institute of Physics, Polish Academy of Sciences, al. Lotnik\'ow 32/46, 02-688 Warszawa, Poland}

\begin{document}
\preprint{May 21, 2010}

\title{Brightening of dark excitons in a single quantum dot containing a single magnetic ion}

\author{M. \surname{Goryca}}\email{Mateusz.Goryca@fuw.edu.pl}\affiliation{\Grenoble}\affiliation{\WHOZA}
\author{P. \surname{P\-lochocka}}\affiliation{\Grenoble}
\author{T. \surname{Kazimierczuk}}\affiliation{\WHOZA}
\author{P. \surname{Wojnar}}\affiliation{\WPAC}
\author{G. \surname{Karczewski}}\affiliation{\WPAC}
\author{J. A. \surname{Gaj}}\affiliation{\WHOZA}
\author{M. \surname{Potemski}}\affiliation{\Grenoble}
\author{P. \surname{Kossacki}}\affiliation{\Grenoble}\affiliation{\WHOZA}

\date{\today}

\begin{abstract}

A promising method to investigate dark exciton transitions in quantum dots is presented. The optical recombination of
the dark exciton is allowed when the exciton state is coupled with an individual magnetic impurity (manganese ion). It
is shown that the efficient radiative recombination is possible when the exchange interaction with the magnetic
ion is accompanied by a mixing of the heavy-light hole states related to an in-plane anisotropy of the quantum dot.
It is also shown that the dark exciton recombination is an efficient channel of manganese spin orientation.

\end{abstract}



\maketitle

Semiconductor quantum dots (QDs) are among the most promising single-photon emitters \cite{Michler_2000_Science,
Michler_2000_Nature, Santori_2001_PRL, Zwiller_2001_APL}. They have potential applications in quantum information
processing, and quantum telecommunications due to their seamless integration in semiconductor circuits, their
robustness, and their relatively easy handling. Crucially, semiconductor QDs provide the possibility to integrate
photonic properties with the spin of an individual magnetic impurity \cite{Besombes_2004_PRL}. The magnetic spin can be selectively manipulated
and used for information storage \cite{LeGaal_2009_PRL, Goryca_2009_PRL2}. However, the use of semiconductor QDs in a
realistic working device requires a reliable control of the excitation process as well as an understanding of the
emission channels.

An important, but nevertheless, little investigated recombination channel is related to the dark exciton states
\emph{i.e.} states with total angular momentum equal to 2 \cite{Cuthbert_1967_PR}. Random transitions between dark and
bright excitonic states lead to exciton decoherence \cite{Palinginis_2004_PRB} and a significant modification of the recombination dynamics which
can result in the delayed emission of photons \cite{Crooker_2003_APL, Labeau_2003_PRL}. Despite their importance, dark exciton states are difficult to probe.
The radiative recombination of dark excitons is forbidden so that they usually cannot be studied directly using
spectroscopic techniques. Their properties can be accessed indirectly by a detailed analysis of the dynamics in
time-resolved profiles of the bright exciton photoluminescence \cite{Bacher_1999_PRL, Crooker_2003_APL, Labeau_2003_PRL}. The other possibility is to
measure the weak optical transitions under conditions in which the dark exciton recombination is partially allowed.
This has been achieved either by the use of the in-plane magnetic field which mixes the heavy-light hole
states \cite{Bayer_2002_PRB, Kowalik_2007_PRB} or by placing the QD in a micro-pillar which enhances the coupling of
the exciton with light \cite{Winger_2008_PRL}.

Here we present an investigation of dark exciton optical transitions which are allowed due to the simultaneous spin
flip of coupled single magnetic impurity. We analyze the dark exciton wave function and show that the radiative recombination of dark excitons is efficient only
when the exchange interaction with the magnetic ion is accompanied by mixing of the heavy-light hole states, related to
an in-plane anisotropy of the QD. To demonstrate the interplay of both mechanisms, high magnetic field spectroscopy has
been employed. We determine all relevant parameters such as the dark exciton oscillator strength, the in-plane
anisotropy, and the exchange interaction. Additionally, we show that the dark exciton recombination can be used as an
efficient channel for controlling the orientation of the spin of the magnetic ion.

The sample, which was grown using molecular beam epitaxy, contains a single layer of self-assembled CdTe QDs with a low
concentration of Mn$^{2+}$ ions, embedded in a ZnTe matrix. The Mn$^{2+}$ concentration was adjusted to obtain a significant number
of QDs containing exactly one Mn$^{2+}$ ion \cite{Wojnar_2007_PRB}. For the measurements, the sample was placed in a
micro-photoluminescence ($\mu$-PL) setup and kept at the temperature of $4.2$~K. The resistive magnet produced magnetic
field up to $28$~T. The field was applied in the Faraday configuration, parallel to the the growth axis of the sample.
The PL of the QDs was excited either above the gap of the ZnTe barrier (at $532$~nm) or using a tunable
dye laser in the range $570-610$~nm. Both the exciting and the collected light were transmitted though a monomode fiber
coupled directly to the microscope objective. The use of the monomode fiber combined with polarization optics outside
the cryostat permits the control of the circular polarization of both the exciting and detected light.

\begin{figure}[t]
\begin{center}
\includegraphics[width=85mm]{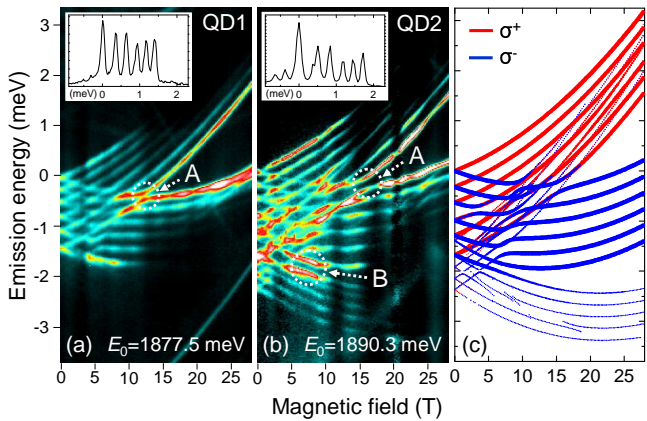}
\end{center}
\caption[]{(a-b) Color-scale plots of the PL spectra of two single Mn-doped QDs as a function of
magnetic field. Two dots (QD1 and QD2) differ by in-plane anisotropy. The vertical scale of each panel is shifted by $E_{0}$. 
Insets: PL spectra at $B=0$. (c) Simulation of the
optical transitions in the QD2 with the model described in text. The line thickness corresponds to the oscillator
strength of the transitions.}
\label{map}
\end{figure}

The diameter of both the excitation and detection spot was less than $2$~$\mu$m, which allows us to select spectra of 
different single QDs containing a single Mn$^{2+}$ ion. Representative results for two
selected QDs with different in-plane anisotropy are shown in the insets of Fig.~\ref{map}. Characteristic PL spectra
contain a neutral exciton ($X$) lines split into sextuplets due to the $X$-Mn exchange interaction
\cite{Besombes_2004_PRL, Fernandez_2006_PRB}. The total spin of the Mn$^{2+}$ ion is $5/2$, and it has six
possible projections onto the direction of highly anisotropic excitonic spin. Thus, each of the six possible spin states of the
ion is related to a specific component of the excitonic sextuplet for a given circular polarization. As in our previous
work \cite{Goryca_2009_PRL2}, we have selected dots containing a single Mn$^{2+}$ ion which have a nonmagnetic QD as a close
neighbor. The resonant excitation of this dot is followed by a spin conserving $X$ transfer to the
Mn-doped QD \cite{Kazimierczuk_2009_PRB}. This permits the efficient and selective excitation of the luminescent dot as 
well as the selection of the spin polarization of excitons injected into this dot.

Figure~\ref{map}(a-b) shows the evolution of neutral exciton PL spectra with magnetic field, measured without
polarization resolution for the two selected QDs. To elucidate the most characteristic spectral features of this evolution we
initially focus on Fig.~\ref{map}(a) showing the QD with the smaller in-plane anisotropy. In the magnetic field, the
exciton sextuplet splits into two distinct Zeeman branches corresponding to $\sigma^{+}$ and  $\sigma^{-}$ circular
polarizations for the upper and lower branch respectively. The lowest component of each branch corresponds to the Mn$^{2+}$
spin antiparallel to the exciton spin \cite{Besombes_2004_PRL}. The resonant excitation with $\sigma^{-}$ polarization
prevents the Mn$^{2+}$ spin from thermalizing in the magnetic field \cite{Goryca_2009_PRL2} due to the interaction of the Mn$^{2+}$
ion with spin-polarized carriers injected into the dot. Therefore all six lines are visible in both branches up to
magnetic fields $\sim 15$~T.

At a magnetic field of around $12$~T, an anticrossing of the outermost lines of both branches is clearly visible (``A''
in Fig~\ref{map}(a)). These two lines correspond to the same $-5/2$ state of the Mn$^{2+}$ spin, but two opposite spin states of
$X$. At this field, the excitonic Zeeman splitting exactly compensates the $X$-Mn exchange interaction. The splitting
(anticrossing) of these two lines is then simply due to the in-plane anisotropy of the QD, acting via the anisotropic
component of the electron-hole (\emph{e}-\emph{h}) exchange interaction, as observed for excitonic lines in the absence of magnetic field
for nonmagnetic QD \cite{Gammon_1996_Science}. As we have checked, the two split lines show linear polarization
(presumably along the symmetry axes of the anisotropic dot), in contrast to the remaining sextuplet lines which are polarized circularly. The
anisotropic exchange splitting, determined from the anticrossing, is equal to $60\mu$eV for QD1 in Fig~\ref{map}(a),
and $230\mu$eV for QD2 in Fig~\ref{map}(b). 

Our optical method to align the Mn$^{2+}$ spin against the action of the 
external magnetic field \cite{Goryca_2009_PRL2} becomes less efficient at magnetic fields above $\sim 15$~T. This is due to the accelerated
spin-lattice relaxation of the Mn$^{2+}$ spin at high magnetic fields \cite{Strutz_1992_PRL}. As a result, the Mn$^{2+}$ spin
orientation thermalizes and the excitonic lines related to the less populated spin states vanish.

Strikingly, the QD with the larger anisotropic exchange splitting value (Fig~\ref{map}(b)) has an additional, albeit
weaker, lower branch consisting of only five lines. To understand the origin of this branch one should notice that
optical transitions are possible in two situations: (a) The projection of total angular momentum of the exciton on the
quantization axis is equal to $\pm1$ (bright exciton). In this case the transition is dipole allowed and the Mn$^{2+}$ spin projection
is conserved during the $X$ recombination. Branches with six lines are related to this kind of recombination.
(b) The projection of total angular momentum of the exciton is equal to $\pm2$ (dark exciton, $X_{d}$). Then, in the first approximation
the optical transition is dipole forbidden. However, the valence band mixing and the exchange interaction with the Mn$^{2+}$
ion result in a mixing of the electron and hole spin states. As a result, the $X_{d}$ states have an admixture of the
$X$ states with the Mn$^{2+}$ spin projection different by $1$. Thus, the $X_{d}$ recombination is possible when accompanied
by the simultaneous spin-flip of the Mn$^{2+}$ ion. As there are only $5$ possible transitions between the $6$ Mn$^{2+}$ spin states,
the PL lines related to $X_{d}$ present a fivefold spitting. The upper energy branch of $X_{d}$ is not clearly visible
in our experiment because it overlaps with much stronger $X$ transitions.

\begin{figure}[t]
\begin{center}
\includegraphics[width=85mm]{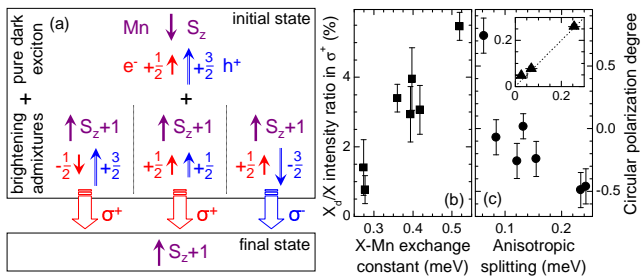}
\end{center}
\caption[]{ (a) Schematic diagram of $X$ admixtures in the $X_{d}$ states and possible channels of optical transitions
related to the $X_{d}$ recombination. (b) $X_{d}$/$X$ intensity ratio in $\sigma^{+}$ polarization \emph{vs.} $X$-Mn
exchange constant. (c) Degree of circular polarization of $X_{d}$ spectrum \emph{vs.} anisotropic exchange splitting of
the QD. A value of $-1$ denotes pure $\sigma^{-}$ polarization, while a value of $+1$ refers to pure $\sigma^{+}$
polarization. Inset: The width of splitting between the $|+1.5, +1/2, +3/2\rangle$ and $|+2.5, +1/2, -3/2\rangle$ lines
(``B'' in Fig.~\ref{map}(b)) calculated from the model \emph{vs.} splitting between these lines determined directly from
the PL for those QDs for which it was possible.}
\label{split}
\end{figure}

The quantum states of the exciton and Mn$^{2+}$ spin can be described in the basis given by three quantum numbers: 
$|S_{z}, \sigma_{z}, j_{z}\rangle$ indicating the Mn$^{2+}$, electron and hole angular momentum projections onto the quantization axis parallel to
the magnetic field. Using the available information for the g-factors of the carriers and the Mn$^{2+}$ ion
\cite{Besombes_2000_JCG}, we attribute the $X_{d}$ low energy branch to the recombination of $|S_{z}, +1/2,
+3/2\rangle$ states. Therefore, to satisfy the selection rules for $X_{d}$ dipolar recombination the projection of the
Mn$^{2+}$ spin must be increased by $1$. This implies that each possible $X_{d}$ recombination is related to a spin-flip of
the Mn$^{2+}$ ion towards the state polarized opposite to the thermalized state. There are three different admixtures of the
$X_{d}$ states which make this process possible (see Fig.~\ref{split}(a)). The first one is the state with the opposite
spin projection of the electron ($|S_{z}+1, -1/2, +3/2\rangle$). It is mixed with the $X_{d}$ state due to the
\emph{e}-Mn exchange interaction. The second admixture, caused by the  \emph{h}-Mn exchange interaction, consists of a light
hole with a spin projection different by $1$ from the heavy hole of the $X_{d}$ ($|S_{z}+1, +1/2, +1/2\rangle$). The
last admixed state, $|S_{z}+1, +1/2, -3/2\rangle$, consists of a heavy hole with a spin projection different by $3$ from
the spin projection of the original state. This admixture is induced by two interactions acting together: the heavy-light hole mixing,
which mixes hole states with spin projection different by $2$, and by the \emph{h}-Mn exchange interaction. A direct
experimental evidence for the presence of valence band mixing is provided by the anticrossing of the highest energy
line of $X_{d}$ and the lowest energy line of $X$ at a magnetic field $\sim7$~T (labeled ``B'' in Fig~\ref{map}(b)).
These lines correspond to $|+1.5, +1/2, +3/2\rangle$ and $|+2.5, +1/2, -3/2\rangle$ states, respectively.

\begin{figure}[t]
\begin{center}
\includegraphics[width=85mm]{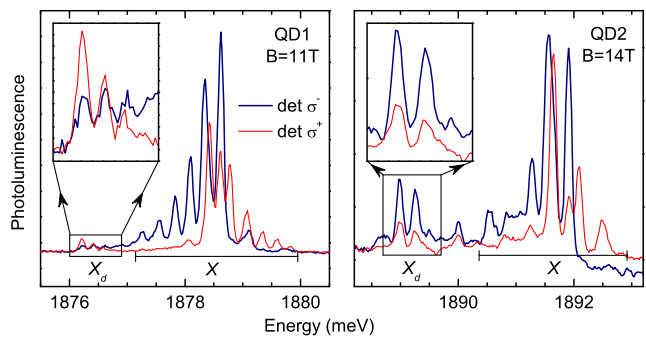}
\end{center}
\caption[]{ Spectra of the two QDs shown in Fig.~\ref{map} taken under excitation with $\sigma^{-}$ polarized light with
polarization resolved ($\sigma^{+}$ or $\sigma^{-}$) detection for magnetic fields near the anticrossing of lines
corresponding to $|-2.5, -1/2, +3/2\rangle$ and $|-2.5, +1/2, -3/2\rangle$ states (``A'' in Fig~\ref{map}(a-b)). }
\label{spectra}
\end{figure}

The radiative recombination related to the first two of these admixtures results in the emission of a photon with
$\sigma^{+}$ polarization. The amplitudes of these admixtures to the $X_{d}$ state depend only on the \emph{e}-Mn and
\emph{h}-Mn exchange constants, \emph{i.e.} on the overlap of the Mn$^{2+}$ ion and the carrier wave functions. In contrast, the
recombination related to the third admixture produces a $\sigma^{-}$ polarized photon. The amplitude of this admixture
depends not only on the $X$-Mn exchange interaction, but also on the valence band mixing, which, similarly to the
anisotropic exchange splitting, results from the in-plane anisotropy of the QD
\cite{Bayer_1999_PRL,Koudinov_2004_PRB,Leger_2007_PRB,Kowalik_2007_PRB}. The role of this anisotropy in determining the polarization
of the $X_{d}$ lines is clearly visible in our experiment. Fig.~\ref{spectra} shows spectra of the two QDs shown in
Fig~\ref{map} for similar excitation conditions and magnetic field near the anticrossing of the lines corresponding to
$|-2.5, -1/2, +3/2\rangle$ and $|-2.5, +1/2, -3/2\rangle$ states (``A'' in Fig~\ref{map}(a-b)). The $\sigma^{-}$
polarized lines are much more pronounced with respect to the $\sigma^{+}$ polarized ones for a highly
anisotropic QD2.

As a quantitative measure of the $X_{d}$ oscillator strength for both circular polarizations we use the $X_{d}$/$X$
intensity ratio. Since this ratio depends on the excitation power \cite{Besombes_2005_PRB}, one should use identical
excitation conditions to be able to compare its value for different QDs. Experimentally, this is achieved by choosing
the same $XX$/$X$ intensity ratio equal in our case to $1/3$. The $XX$ intensity increases roughly quadratically with
the excitation power, while the $X$ intensity follows linear power dependence (\cite{Brunner_1994_PRL,
Suffczynski_2006_PRB}). Thus, this ratio gives a measure of the excitation efficiency.

A strong increase of the $X_{d}$/$X$ intensity ratio in $\sigma^{+}$ polarization with increasing $X$-Mn exchange
constant is clearly visible in Fig.~\ref{split}(b). It confirms the origin of the first two recombination channels
described above. However, the $X_{d}$ oscillator strength for $\sigma^{-}$ polarization depends both on the $X$-Mn
exchange interaction and the QD anisotropy. To elucidate the role of the anisotropy, we can use the circular polarization of the $X_{d}$ . Fig.~\ref{split}(c) shows that the polarization of the $X_{d}$ evolves towards $\sigma^{-}$
as the anisotropic exchange splitting increases. Therefore, we conclude that a high in-plane anisotropy of the QD leads
to a strong mixing of light and heavy holes, which  combined with the presence of the Mn$^{2+}$ ion is the reason for the
strong $\sigma^{-}$ polarization and high intensity of the $X_{d}$ spectrum. It is important to note that for
nonmagnetic QDs showing similar anisotropy \cite{Kowalik_2007_PRB} the optical transitions of $X_{d}$ are not observed,
except in the presence of a strong in-plane magnetic field.

As each $X_{d}$ recombination in the lower energy branch involves an increase of the Mn$^{2+}$ spin projection by $1$, dark
exciton recombination can play the role of an effective Mn$^{2+}$ spin orientation mechanism. In our experiment the
$X_{d}$/$X$ intensity ratio for a highly anisotropic QD and $\sigma^{-}$ polarized excitation was as high as 10\%. This
is comparable to the probability of a spin-flip of the Mn$^{2+}$ ion per one recombination of the $X$ in the QD which was
estimated to be $\sim10$\% in Ref.~\cite{Goryca_2009_PRL2}. While this orientation mechanism should be present for both
circular polarizations of excitation, only the $\sigma^{-}$ polarization which populates the low energy branch of $X_{d}$ is fully seen.
The $\sigma^{+}$ polarized excitation decreases the Mn$^{2+}$ spin projection \cite{Goryca_2009_PRL2}. Under such conditions the high energy branch of 
$X_{d}$ should be populated and the low energy branch should be virtually invisible. The latter is, indeed, confirmed in our experiment. 
However, it is not possible to observe directly the high energy $X_{d}$ branch, since it occurs in the same energy region 
as the much stronger $X$ lines.

A quantitative description of key features of the data in Fig.~\ref{map}(a-b) is provided by a simple model with the
initial state of the QD after excitation event given by the following Hamiltonian
(\cite{Kesteren_1990_PRB,Bayer_2002_PRB, Ivchenko_1996_PRB, Toropov_2000_PRB, Kowalik_2007_PRB}),
\begin{eqnarray*}
\label{eq:eq1}
\mathcal{H} = g_{Mn}\mu_B\overrightarrow{B}.\overrightarrow{S} + g_{e}\mu_B\overrightarrow{B}.\overrightarrow{\sigma} + g_{h}\mu_{B}\overrightarrow{B}.\overrightarrow{j} - I_{e}\overrightarrow{S}.\overrightarrow{\sigma} \\
{} - I_{h}\overrightarrow{S}.\overrightarrow{j} + \sum_{i=x,y,z}\left(a_{i}j_{i}\sigma_{i} + b_{i}j_{i}^{3}\sigma_{i}\right) - \gamma j_{z}^{2} + \beta\left(j_{x}^{2}-j_{y}^{2}\right)
\end{eqnarray*}
where $S$, $\sigma$ and $j$ are the Mn$^{2+}$, electron and hole spin operators, respectively, the first three terms represent
the Zeeman energy of the Mn$^{2+}$ ion, the electron and the hole, $I_e$ and $I_h$ are the \emph{e}-Mn and \emph{h}-Mn exchange interaction
constants, $a_{i}$ and $b_{i}$ are \emph{e}-\emph{h }spin-spin coupling constants, $2\gamma$ is the heavy-light hole splitting and
$\beta$ represents the strength of the heavy-light hole mixing. The first term is also the Hamiltonian of the final
state of the system after the exciton recombination. We also introduced an additional, phenomenological term related to
the excitonic diamagnetic shift to facilitate a comparison of the model and experimental data.

The energies of the optical transitions versus magnetic field, calculated using this Hamiltonian for the two circular
polarizations, are plotted in Fig.~\ref{map}(c). The calculations clearly reproduce the key features of the
experimental data in Fig.~\ref{map}(a-b), such as for example the $X$-Mn exchange splitting and the anisotropic
exchange splitting. All parameters in the Hamiltonian (except for $\gamma$ assumed to be $15$~meV \cite{Leger_2007_PRB}) can be extracted by fitting to the experimental data. In
particular, the heavy-light hole mixing can be estimated using the degree of circular polarization of $X_{d}$ lines.
Such an approach permits an estimation of the $\beta$ parameter even for those QDs, for which the anticrossing between
the $|+1.5, +1/2, +3/2\rangle$ and $|+2.5, +1/2, -3/2\rangle$ lines (``B'' in Fig.~\ref{map}(b)) is not clearly
visible. As shown in the inset of Fig.~\ref{split}(c) the obtained value of this anticrossing remains in very good
agreement with the value estimated directly from PL data. This confirms the proposed mechanism of $X_{d}$
brightening.

To conclude, we have used a QD with a single magnetic impurity (Mn$^{2+}$) to investigate the dark exciton transitions. The
$X$-Mn exchange interaction, when combined with a mixing of
the heavy-light hole states induced by the QD in-plane anisotropy, allows dark exciton recombination 
accompanied by a simultaneous Mn$^{2+}$ spin flip. High magnetic fields have been used to spectrally separate the PL
lines related to $X$ and $X_{d}$ transitions and to extract the important QD parameters (\emph{e.g.} the anisotropic
exchange splitting). The excitation of the Mn-doped QD via a closely lying second QD
permits the precise control of the spin of the carriers injected to the QD and therefore a control over the Mn$^{2+}$ spin
orientation. The simple Hamiltonian, which describes this system, reproduces correctly all the key features of the PL
spectra of the QD in high magnetic field.

\begin{acknowledgments}
This work was supported by the Polish Ministry of Science and Higher Education as research grants in years
2006-2011, by the 6th Research Framework Programme of EU (contract MTKD-CT-2005-029671), by the CNRS PICS-4340
Programme and by the Foundation for Polish Science. Two of us (P.P. and P.K.) are supported by the EU under FP7, 
contracts no. 221249 `SESAM' and no. 221515 `MOCNA' respectively. We thank Anna Trojnar and Marek Korkusi\'nski for
fruitful discussions.
\end{acknowledgments}

\bibliographystyle{apsrev4-1}

\end{document}